\documentstyle[seceq,epsf,wrapft,twoside]{ptptex}
\newcommand{\delslash}{\partial \hspace{-6pt} /}
\newcommand{\sigvev}{\sigma_{0}}
\newcommand{\eqref}[1]{(\ref{#1})}
\newcommand{\zerohalf}{(0,\textstyle \frac{1}{2})}
\newcommand{\halfzero}{(\textstyle \frac{1}{2},0)}
\newcommand{\onehalf}{(1, \textstyle \frac{1}{2})}
\newcommand{\halfone}{(\textstyle \frac{1}{2},1)}
\newcommand{\nstar}{N^{*}}

\notypesetlogo  
\title{Chiral doublet model for positive and negative parity nucleons
}
\author{%
D. Jido\footnote{present address: ECT*,
European Centre for Theoretical Studies in Nuclear Physics and Related Areas, 
Villa Tambosi, Strada delle Tabarelle 286,  I-38050 Villazzano (Trento), Italy.
}
}
\inst{%
Research Center for Nuclear Physics (RCNP), Osaka University \\
Ibaraki Osaka 567-0047, Japan
}
\abst{%
We discuss the properties of the nucleon ($N$) and its excited state
with odd parity ($N^{*}$) under the aspect of chiral symmetry, identifying
them with suitable representations of the chiral group.
It is shown that there are two distinctive schemes to assign
the chiral multiplet to $N$ and $N^{*}$.
We construct linear $\sigma$ models for $N$ and $N^{*}$
with the two assignments to show their physical implications. 
We also discuss the in-medium properties of $N(1535)$, comparing
the present model with the other chiral model
based on the different picture of $N^{*}$.
}

\begin{document}
\maketitle

\setcounter{tocdepth}{4}

\section{Introduction}

Chiral symmetry (ChS) is one of the key concepts for
understanding the structure and dynamics of hadrons at low energies 
from the viewpoint of QCD. The importance of the 
dynamical breakdown of ChS is summarized in the
existence of the light pion as a Nambu-Goldstone boson. 
On the other hand, we believe restoration of the broken symmetry
at hot and/or dense matter, and
one of the interesting implications there is the appearance of 
the degenerate spectra in parity partners.
It has been recently pointed out that partial restoration of ChS
may take place in a nucleus and it will be observed as
effective modifications of hadron properties in the nuclear medium\cite{kunihiro}.

The $N(1535)$ ($N^{*}$) is an especially interesting nucleon resonance.
It is the first excited state with odd parity and is considered 
to be a possible candidate of a chiral partner of the nucleon.
In addition, it is well known that creation of $N^{*}$ in intermediate states
is identified by emission of the $\eta$ meson in the final state
as a result of the strong $\eta NN^{*}$ coupling. 
The recente theoretical studies of $N^{*}$  respecting ChS
have been done in two distinct pictures of $N^{*}$: (1) a chiral partner of the 
nucleon \cite{DeTar:1989kn,Jido:1998yk,Jido:1998av,Jido:2001nt},
(2) a dynamically generated object
in meson-baryon scattering\cite{Kaiser:1995cy,Inoue:2001ip}.

First of all, it is worth emphasizing that 
ChS is unambiguously defined
in the QCD Lagrangian for the quark field as separated rotations in
the flavor space:
\begin{equation}
   q_{l} \rightarrow L q_{l}  \hspace{1cm} q_{r} \rightarrow R q_{r} \ ,
\end{equation}
where $q_{l}$ ($q_{r}$) is the left- (right-) handed component of
the quark field in the sense of Lorentz group, 
and $L$ and $R$ are independent $SU(N_{f})$ rotations for the $N_{f}$ flavors.

On the other hand, the realization of ChS in hadrons is not 
trivial issue as the reflections of their quark composite structure and 
the spontaneous breaking of ChS.
There are two possible ways of the realization of ChS for hadrons,
non-linear and linear realizations. 
They are not conflicting concepts but compliment each other.

In the nonlinear realization, the effective Lagrangians are constructed on the
premise that ChS is spontaneously broken.
The manifestation of ChS in Lagrangian is accomplished
by giving the special role 
in the axial transformation to the Nambu-Goldstone boson.
The transformation rule of the other hadron under ChS is uniquely 
fixed, once the transformation rule under 
the vector rotation
is given~\cite{Weinberg:1968de}.
Dynamical properties are determined according to an expansion 
in powers of momenta of Nambu-Goldstone bosons. This leads us to obtain
the most general Lagrangian at low energies, which is summarized 
in chiral perturbation theory \cite{Weinberg:1979kz,Gasser:1984yg}.

On the other hand, in the linear representation, based on
the fact that all hadrons are in principle classified into some representations
of the chiral group $SU(N_{f})_{L}\otimes SU(N_{f})_{R}$, 
the effective Lagrangians are constructed with assigning irreducible
representations to the hadrons. The linear realization has a connection
between chiral symmetric and broken phase,
and gives constraints on intrinsic properties of hadron 
appearing in the broken phase, such as the mass of the nucleon.

In this paper, we would like to discuss the nucleon and its excited state with
odd parity in a unified formulation based on the linear
realization, considering the fact that the parity degeneracy appears 
in the restoration limit of ChS, 
where the nonlinear realization breaks down. 
We discuss the lowest-lying $N(1535)$ as the excited state 
with odd parity considered here, although
it can be, in principle, assumed to be any nucleon excited state 
with $J^{\pi}=(1/2)^{-}$.

In Sec.\ 2, we discuss how to realize chiral symmetry for $N$ and $N^{*}$
in the linear representation and see that there are two possible ways to
assign the chiral transformation to $N^{*}$. 
According to these assignments, in Sec.\ 3,
we construct the two chiral doublet models and discuss their physical
consequences.
In Sec.\ 4, we apply the chiral double model to the study of 
the in-medium properties of $N(1535)$ probed 
by $\eta$ mesic nuclei. Summary is presented in Sec.\ 5.

\section{Chiral symmetry for baryons}

In order to construct an effective Lagrangian in the linear representation,
it is necessary to assign an appropriate irreducible representation
of the chiral group to the nucleon $N$ and its odd parity excited state $N^{*}$.
The chiral multiplet is a good quantum number in the chiral symmetric limit
and represents the quark configuration inside the nucleons.
The most suitable combinations of the chiral multiplets for the nucleon
should be in principle determined by the dynamics of the quarks and gluons. 
Now let us concentrate the flavor two case ($N_{f}=2$) 
and the chiral limit $m_{q}=0$, and we do not consider 
the possible mixing to the other representations.

As mentioned in introduction, ChS in QCD is defined 
in terms of the quark field, and the $u$- and $d$-quark fields belong to
the fundamental representation
$\halfzero\oplus \zerohalf $ where the first and second numbers 
in the parenthesis expresses the irreducible representations 
of $SU(2)_{L}$ and $SU(2)_{R}$, respectively.
Considering that the baryons consist of three valence quarks, 
possible candidates of the chiral multiplet for the baryons may be given by
the following three multiplets\cite{Jido:1998yk,Cohen:1997sb}:
\begin{equation}
  \left[ \halfzero \oplus \zerohalf\right]^{3}
  =  5 \left[ \halfzero \oplus \zerohalf\right] \oplus
  3  \left[ \halfone \oplus \onehalf\right] \oplus
  \left[\left(\textstyle\frac{3}{2},0\right) \oplus 
  \left(0, \textstyle\frac{3}{2}\right) \right]  
\end{equation}
The terms in the first and third parentheses in r.h.s.\ 
have purely isospin $1/2$ and $3/2$, respectively, while the terms
in the second parenthesis is the mixture representation of the isospin
$1/2$ and $3/2$. Here we take for $N$
the first representation  $\halfzero\oplus \zerohalf$, which has the isospin 
$1/2$.
This representation has been used in the linear $\sigma$ model by Gell-Mann 
and L\'evy, 
the QCD sum rules\cite{Ioffe:1981kw} and lattice QCD calculations.
We assume that the equivalent multiplet to the nucleon is assigned to
$N^{*}$.

After the choice of the irreducible representation for both nucleons,
there are two possible way to assign the chiral multiplet to $N$ and $N^{*}$,
depending on how to introduce the mass terms
of the nucleons consistently with ChS. 

In the first case, the transformation rules for $N$ and $N^{*}$ are given by
\begin{equation}
\begin{array}{cc}
      N_{l} \rightarrow L N_{l}\ \ \ & \ \ \  N_{r} \rightarrow R N_{r}\\
      N_{l}^{*} \rightarrow L N_{l}^{*}\ \ \ & \ \ \  
      N_{r}^{*} \rightarrow R N_{r}^{*}
\end{array}
  \ , \label{tranai}
\end{equation}
which is called as the {\it naive model} \cite{Jido:1998av}.
The matrices $L$ and $R$ in \eqref{tranai} represent the rotations of 
$SU(2)_{L}$ and $SU(2)_{R}$, respectively,  and, 
$N_{l,r}$ and $N_{l,r}^{*}$ are their doublets. 
In this case, the explicit introduction of the nucleon mass terms 
breaks invariance under the axial transformation of ChS.
The only prescription to make models invariant under ChS
is that the nucleons is introduced as massless
Dirac particles coupling to a scalar field and condensate of the scalar field 
induces generation of the nucleon masses as well as 
the spontaneous symmetry breaking. 
However, in this case, the two nucleons belong to the completely 
separate multiplet and transform independently under ChS.
Therefore there are no connections between them
in the group theoretical point of view. If $N^{*}$ were assigned to even
parity, there would be no changes in 
the argument presented here for the naive model because of no
connection between $N$ and $N^{*}$ in terms of ChS.

Alternatively it is possible to keep the invariance of the mass term under
the linear transformation when we introduce the chiral partner 
of the nucleon, which is the particle to form the parity degeneracy in 
the restoration limit of ChS. 
Let us consider the nucleons $N$ and $N^{*}$ which are the chiral partners and
transform each other under the axial transformation of ChS,
similarly to the $\sigma$ and $\pi$ fields. 
The nucleon mass term is written with a common mass $m_{0}$ in a
chiral invariant way as
\begin{equation}
  m_{0} (\bar NN + \bar N^{*}N^{*}) \label{invmass} \ ,
\end{equation}

The physical basis $(N, N^{*})$ is different form the 
basis of ChS, which is defined in the transformation rule 
under $SU(2)_{L}\otimes SU(2)_{R}$ as
\begin{equation}
\begin{array}{ll}
  N_{1r} \rightarrow R N_{1r} \ \ \ &\ \ \  
  N_{1l} \rightarrow L N_{1l}\\
  N_{2r} \rightarrow L N_{2r}\ \ \  &\ \ \ 
  N_{2l} \rightarrow R N_{2l} \label{tramir}
 \end{array} \ ,
\end{equation}
where $N_{1}$ and $N_{2}$ are assumed to have even
and odd parity, respectively.
Note that $N_{1}$ and $N_{2}$ transform in the opposite way under
the axial transformation. This is called as 
the {\it mirror model} \cite{Jido:1998av}.
It is possible to introduce the following mass term without
any contradictions with ChS\cite{DeTar:1989kn}:
\begin{equation}
   m_{0} \left(\bar N_{1} \gamma_{5} N_{2}-
    \bar N_{2} \gamma_{5} N_{1}\right) \label{mirmassterm} \ .
\end{equation}
The physical basis is obtained so as to diagonalize 
the mass term \eqref{mirmassterm}:
\begin{equation}
  N = \textstyle\frac{1}{\sqrt 2} \left(N_{1} + \gamma_{5} N_{2}\right) \ \ \ \ \ \ 
  N^{*} = \frac{1}{\sqrt 2} \left(\gamma_{5} N_{1} - N_{2}\right)   \ .
\end{equation}
It is shown that $N$ transforms to $N^{*}$ under the axial 
transformation \eqref{tramir}.

Here let us make a remark regarding to the axial $U(1)$ symmetry. 
The QCD Lagrangian is invariant under the global  $U(1)$ axial transformation,
and the $U(1)_{A}$ symmetry is anomalously broken due to the quantum
effect. In the present work, although we are constructing the effective models 
for hadrons which have the same symmetries as QCD, we do not assume the
$U(1)_{A}$ symmetry in the effective models, since the effective models
emerge after the quark loops are integrated out. Nevertheless, the $U(1)_{A}$
charge of hadron gives further constraints on the quark structure of the hadron
\cite{Christos:1983kc}. For instance,
the chiral multiplet  $\halfzero\oplus \zerohalf$ considered here is composed
in the two different ways \cite{Cohen:1997sb}, $q_{l}(q_{l}q_{l})_{I=0} 
\oplus q_{r}(q_{r}q_{r})_{I=0}$
and $q_{l}(q_{r}q_{r})_{I=0} \oplus q_{r}(q_{l}q_{l})_{I=0}$. Both are
the same multiplet in the $SU(2)_{L}\otimes SU(2)_{R}$ group, but have
the different charges of $U(1)_{A}$.

\section{Chiral double models}
In this section we briefly discuss the phenomenological consequences of 
the two assignments introduced in the preceding section, constructing
the linear $\sigma$ models according to their transformation rules. The detailed
discussion has been shown in Refs.~\citen{Jido:1998av,Jido:2001nt}.
The important consequences are summarized in Table 1.

Considering the chiral transformation rule for the scalar and pseudo-scalar fields
$M\equiv\sigma + i \vec\tau \cdot \vec\pi \rightarrow L M R^{\dagger}$, 
we obtain the linear $\sigma$ model with the naive assignment as
\begin{equation}
  {\cal L}_{\rm naive} = \sum_{j=1,2}\left[ \bar N_{j} i \delslash N_{j}
   - a_{j} \bar N_{j} (\sigma + i\gamma_{5} \vec\tau \cdot \vec\pi) N_{j} \right]
   + {\cal L}_{\rm meson}
   \label{eq:lagnai} \ ,
\end{equation}
where $N_{1} \equiv N$ and $N_{2}\equiv N^{*}$, and $a_{1}$ and $a_{2}$
are free parameters independent of ChS.
The Lagrangian ${\cal L}_{\rm meson}$ is for the $\sigma$ and $\pi$ fields,
and its explicit form is irrelevant for the present
argument as long as it causes the spontaneous chiral symmetry breaking. 
The transition coupling of $N$ and $N^{*}$ with the meson field is also
possible to be introduced 
in the chiral invariant way\cite{Jido:1998av,Jido:2001nt},
but the coupling term is always diagonalized by an suitable 
linear combination of $N$ and $N^{*}$ without any contradiction 
with the chiral transformation rule \eqref{tranai}. 
This Lagrangian has the minimal terms invariant under ChS and
it is allowed to add more terms with derivatives and multiple $M$
fields. These terms give corrections to the axial charges and the masses 
of the nucleons 
in powers of the sigma condensation $\sigvev$. 

The Lagrangian of the naive model is just a 
sum of two independent linear $\sigma$ models. 
Therefore the phenomenological consequences are followed by those
of the usual linear $\sigma$ model for the single nucleon. 
The masses of $N$ and $N^{*}$ are calculated with the finite condesate
of the scalar field as
\begin{equation}
  m_{N} = a_{1} \sigvev \ , \ \ \ \ \ \ \ m_{N^{*}} = a_{2} \sigvev \ .
  \label{naimass}
\end{equation}
The isovector axial charges of $N$ and $N^{*}$ are unities 
independently of $\sigvev$ at tree level.
There are no transitions between $N$ and $N^{*}$ with pion, which 
is qualitatively consistent with the empirically small value of 
the $\pi NN^{*}$ coupling $g_{\pi NN^{*}}\simeq 0.7$ compared to 
the strong $\pi NN$ coupling $g_{\pi NN}\simeq 13$. 

Another interesting consequence 
is that the values of the masses and the axial charges in
the chiral restoration limit; $m_{N}= m_{N^{*}} =0 $, 
$g_{A}^{N}= g_{A}^{N^{*}}=1$ and $g_{A}^{NN^{*}}=0$.
This is a quite general conclusion.
Even if we add more terms invariant under ChS to 
Lagrangian \eqref{eq:lagnai}, their contributions to 
the masses and the axial charges are written in powers of $\sigvev$,
and, therefore, the masses of $N$ and $N^{*}$ are 
decreasing to zero and their axial charges are 
approaching to unity and zero at least close to 
the restoration limit\cite{Kim:1998up}.

Now let us turn to the mirror model. The Lagrangian is given by
\begin{eqnarray}
  {\cal L}_{\rm mirror}& =& \sum_{j=1,2}\left[ \bar N_{j} i \delslash N_{j}
   - g_{j} \bar N_{j} (\sigma + (-)^{j+1} i\gamma_{5} \vec\tau \cdot \vec\pi) N_{j} 
   \right] \nonumber \\
   &&   - m_{0} \left( \bar N_{1}\gamma_{5} N_{2}
           - \bar N_{2}\gamma_{5} N_{1}\right)  + {\cal L}_{\rm meson}
           \label{eq:lagmir} \ ,
\end{eqnarray}
where $g_{1}$, $g_{2}$ and $m_{0}$ are free parameters, and we assume
to truncate the terms with derivatives and multi mesons again. The Yukawa 
term mixing $N_{1}$ and $N_{2}$ is not invariant under the 
transformation~\eqref{tramir}.
This Lagrangian was first formulated and investigated by 
DeTar and Kunihiro~\cite{DeTar:1989kn}. Historically  
a similar Lagrangian to the present one was considered before
by B.~Lee in~\citen{LeeBook1972}, but symmetry 
between $N$ and $N^{*}$ under the axial $U(1)$ transformation
was implicitly assumed in his Lagrangian 
and he got physically uninteresting results.

As mentioned in the previous section, the physical nucleons $N$ and $N^{*}$
diagonalizing the mass term are given by a linear combination of 
$N_{1}$ and $N_{2}$. After breaking ChS spontaneously, $N$ and $N^{*}$
are given with a mixing angle $\theta$ as
\begin{equation}
N =  \cos\theta  N_{1} + \sin\theta \gamma_{5} N_{2}\ ,
\ \ \ \ \ \ \
N^{*} =   -\sin\theta  \gamma_{5}N_{1} +  \cos\theta N_{2} \ .
\end{equation}
The mixing angle depends on the sigma condensate:
$
    \tan 2\theta = {2m_{0}}/{\sigvev (g_{1} + g_{2})}
$.
The corresponding masses are calculated as
\begin{equation}
   m_{N,N^{*}} = \frac{1}{2} \left( \sqrt{(g_{1}+g_{2})^{2} \sigvev^{2}
   + 4 m_{0}^{2}} \mp (g_{2}-g_{1})\sigvev\right) \ . \label{mirmass}
\end{equation}
In this model, 
mass degeneracy of $N$ and $N^{*}$ takes place with a finite mass $m_{0}$
in the chiral restoration limit, and the spontaneous breaking of ChS causes
the mass splitting.  
The model parameters are fitted so as to reproduce the $N$ and $N^{*}$ masses
and the $\pi NN^{*}$ coupling: $g_{1}=9.8$, $g_{2}=16$ and $m_{0}=270$ MeV
\cite{DeTar:1989kn,Jido:1998av,Jido:2001nt}.
The mixing angle is calculated as $\theta = 6.3^{\circ}$. The similar mass formula
to \eqref{mirmass} has been obtained for the isospin $1/2$ and $3/2$ nucleon 
resonances, such as $\Delta$, $N(1520)$ and their parity partners, in the same
approach presented here with the multiplet $(\halfone\oplus\onehalf)$, 
and it is
consistent with the observed pattern of the mass spectra \cite{Jido:1999hd}. 

One of the phenomenological significances of this model is that the 
axial charges of $N$ and $N^{*}$ have the opposite sign to each
other. The isovector axial charges are calculated in a function
of the mixing angle as 
\begin{equation}
    g_{A} = \left(
    \begin{array}{cc} \cos 2\theta &  - \sin 2 \theta \\ 
     -\sin 2\theta & -\cos 2\theta \end{array}\right) \ .
\end{equation}
This shows that the relative sign of the axial charges of $N$ and $N^{*}$ 
is negative independently of the mixing angle
and that the off-diagonal component does not necessarily vanish. It is 
also shown that in the 
chiral restoration limit ($\theta = \pi/4$) the diagonal components
are zero, while the off-diagonal component is unity. Considering 
the empirical value of the transition axial charge 
$g_{A}^{NN^{*}}\simeq 0.2$ obtained by the generalized
Goldberger-Treiman relation with $g_{\pi NN^{*}} \simeq 0.7$,
the transition axial charge $g_{A}^{NN^{*}}$ is enhanced to unity as the
sigma condensate decreases\cite{Kim:1998up}. 

The parameter $m_{0}$ introduced here is a new parameter 
not constrained by ChS, and it gives the nucleon mass 
in the chiral restoration limit. If the mirror model is realized in the physical 
nucleon, the origin of $m_{0}$ in QCD is important 
to understand the mirror nucleons in the 
context of non perturbative QCD\cite{Jido:2001nt}.  

\begin{table}[thb]
\caption{Summary of the phenomenological consequences of the naive and
mirror models.}
\label{summary}
\begin{tabular}{|c|cc|}
\hline
 & Naive model & Mirror model\\
 \hline
Definition & 
    $\begin{array}{cc}
           N_{1r} \rightarrow R N_{1r}, & N_{1l} \rightarrow L N_{1l} \\
           N_{2r} \rightarrow R N_{2r}, & N_{2l} \rightarrow L N_{2l}
    \end{array} $ &
    $\begin{array}{cc}
           N_{1r} \rightarrow R N_{1r}, & N_{1l} \rightarrow L N_{1l} \\
           N_{2r} \rightarrow L N_{2r}, & N_{2l} \rightarrow R N_{2l}
    \end{array} $\\
Nucleon mass term & generated by scalar field & introduced with chiral partner\\
Nucleon in Wigner phase &  massless & massive\\
Role of $\sigvev$ & mass generation & mass splitting \\
Chiral partner &    $\begin{array}{cc}
           N \leftrightarrow \gamma_{5} N  &
           N^{*} \leftrightarrow \gamma_{5} N^{*} 
    \end{array} $ & $N \leftrightarrow N^{*}$\\
Sign of $\pi N^{*}N^{*}$ coupling & positive & negative\\
In-medium $\pi NN^{*}$ coupling& suppressed & enhanced \\
\hline     
\end{tabular}
\end{table}

\section{Application: {\it in-medium properties of $N(1535)$ probed 
by $\eta$ mesic nuclei}}
In this section, we show an application of the parity doublet model to
the investigation of the  in-medium properties of $N(1535)$.
It has been pointed out in Ref. \citen{Jido:2002yb} 
that formation experiments of $\eta$ mesic nuclei, 
such as ($d,^{3}$He) reactions with nuclear targets, 
could be good tools to observe the in-medium effect on the mass of $\nstar$.

The basic idea is that the $\eta$ optical potential in the nucleus is expected 
to be very sensitive to medium modifications of the $\nstar$ mass. Assuming
the $\nstar$ dominance hypothesis and the heavy baryon limit, we obtain
the $\eta$ optical potential as
\begin{equation}
  V_{\eta} = \frac{g_{\eta}^{2}}{2 \mu} \frac{\rho(r)}{\omega + 
  m^{*}_{N}(\rho) - m^{*}_{N^{*}}(\rho) + i\Gamma_{N^{*}}(s; \rho)/2}
\end{equation} 
with $g_{\eta}$ the $\eta NN^{*}$ coupling, $\mu$ the reduced mass
of the $\eta$-nucleus system and $\rho(r)$ the nuclear density.
Considering the fact that the $\nstar$ mass in free space lies only 50 MeV 
above the $\eta N$ threshold, we conclude that the $\eta$-nucleus potential
turns to be repulsive at the nuclear center, if the in-medium effect leads to a
significant reduction of the mass difference of 
$N$ and $\nstar$\cite{Jido:2002yb,Jido:2002mg,Nagahiro:2003iv}.
We expect to observe the repulsive nature in the formation experiment 
of  $\eta$ mesic nuclei with the ($d,^{3}$He) reactions.

We use the parity doublet model to calculate the $N$ and $N^{*}$ masses
and the $N^{*}$ decay width in medium. We assume the partial restoration 
of ChS in nuclei with a linear parametrization of the density
dependence of the sigma condensate\cite{Hatsuda:1999kd,Jido:2000bw}:
\begin{equation}
  \langle \sigma \rangle = (1-C \rho/\rho_{0}) \sigvev \label{sigmed}
\end{equation}
with $C=0.1$ - $0.3$ and $\rho_{0}$ the saturation density. 
Since the mass difference of $N$ and $N^{*}$ is proportional to 
the sigma condensate in both chiral doublet models as seen
in eqs.\ \eqref{naimass} and \eqref{mirmass}, the in-medium 
mass difference is obtained as
\begin{equation}
   m_{N^{*}}^{*}(\rho) - m_{N}^{*}(\rho)
    = (1-C \rho/\rho_{0}) (m_{N^{*}} - m_{N})
\end{equation}
with the density dependent sigma condensate \eqref{sigmed}.
Therefore, as long as partial restoration of ChS is 
assumed in the nuclear medium, 
the mass difference is reduced as the density increases and 
the $\eta$ optical potential in nucleus
has possibility to turn to be repulsive at the center. 

We show in Fig.~1 
the spectra of the ($d,^{3}$He) reaction with $^{12}$C target
calculated in the various cases of the $\eta$-nucleus interaction.
In Fig.~1~(a), the medium modifications of the $N$ and $N^{*}$ masses
are not assumed
and the $\eta$ optical potential is attractive in the nucleus independently
of the density. 
This case is equivalent to the so-called $T$-$\rho$ approximation.
Shown in Fig.~1~(b) is the spectrum calculated in the mirror model
with the partial restoration of ChS and 
its strength parameter $C=0.2$. Here we find the significant difference in 
these two plots. 
In the mirror model, as a result of the repulsive nature
of the $\eta$ optical potential at the center of nucleus, the spectrum
is shifted to the higher energies. 
Note that the peak structure shown in the plots is not responsible for
the formation of the $\eta$ bound state but just the 
threshold effects\cite{Nagahiro:2003iv}.

It is also interesting to compare the above results with the spectrum 
calculated by the chiral model based on the different picture of $N^{*}$, 
such as a dynamically generated 
object in meson-baryon scattering. This model was formulated first in 
Ref.~\citen{Kaiser:1995cy}. There
$\nstar$ is calculated in the coupled channels 
of $\pi N$, $\eta N$, $K \Lambda$ and $K \Sigma$, and the
$\nstar$ is found to be formed dominantly 
as a $K\Sigma$ state~\cite{Kaiser:1995cy}. Since
the in-medium modification of $\nstar$ is 
insignificant in this model\cite{Waas:1997pe,Inoue:2002xw},
the optical potential of $\eta$ in nuclei is 
basically attractive inside of the nuclei. 
Here we directly take the in-medium $\eta$ optical potential
shown in Ref.~\citen{Inoue:2002xw} to calculate the ($d,^{3}$,He) spectrum. 
As shown in Fig.~1~(c), as a result of the irrelevance of the in-medium modification 
of $\nstar$, the spectrum has the similar shape to Fig.~1~(a).
We would expect that the spectra obtained with the different pictures
of $N^{*}$ are distinguished in experiment.

\begin{figure}[thb]
\epsfxsize=13cm
\centerline{
\epsfbox{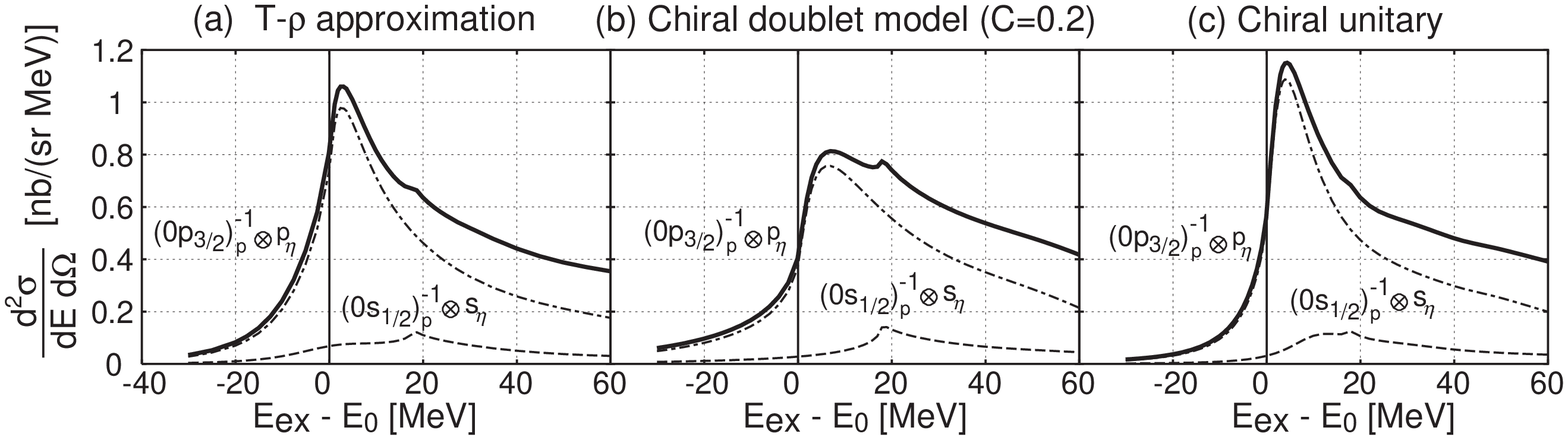}}
\caption{The calculated spectra of $^{12}$C($d,^3$He)$^{11}$B$\otimes\eta$
reaction at $T_d$=3.5 GeV are shown as functions of the excited energy
$E_{{\rm ex}}$. $E_0$ is the $\eta$ production
threshold energy. The $\eta$-nucleus interaction is calculated by
(a) the $t \rho$ approximation, (b) the chiral doublet model with
$C=0.2$ and (c) the chiral unitary approach.  The total spectra are
shown by the thick solid lines, and
the dominant contributions
from the $(0s_{1/2})^{-1}_p \otimes s_\eta$ and
the $(0p_{3/2})^{-1}_p \otimes p_\eta$ configurations are shown by
dashed lines and dash-dotted lines, respectively. Here the proton-hole states
are indecated as $(n\ell_{j})^{-1}_{p}$ and the $\eta$ state as $\ell_{\eta}$.
\label{fig:12C_target2}}
\end{figure}

\section{Summary}
We have investigated the properties of the nucleon and its excited state
with odd parity in the effective models which are strongly constrained
by chiral symmetry. There are two possible ways to assign the chiral
transformation to $N^{*}$ in the linear realization of chiral symmetry.
So far we do not know which models is realized in the physical nucleon
and excited state. To confirm it experimentally, the important observable
is the sign of the isovector axial charge of $N^{*}$ \cite{Jido:2000nt,Jido:2001nt}.

We have also discussed the in-medium properties of $N(1535)$ probed
by $\eta$ mesic nuclei based on the two distinct physical pictures of
the structure of  $N^{*}$.
We have found that the models based on these pictures
produce quantitatively different consequences and 
they would be distinguishable in formation experiments of $\eta$-mesic nuclei,
for instance, the ($d,^{3}$He) reaction with nuclear target. 

\acknowledgements
I would like to express  my sincere gratitude to Prof.\ M.\ Oka, Prof.\
A.\ Hosaka, Dr.\ Y.\ Nemoto, Dr.\ H.\ Kim, Prof.\ T.\ Hatsuda, 
Prof.\ T.\ Kunihiro, Prof.\ S.\ Hirenzaki and Dr.\ H.\ Nagahiro, 
who have participated in the works reported here. 


\begin{thebibliography}{10}
\bibitem{kunihiro}
See, for example, T.~Kunihiro, the contribution to this conference proceedings. 

\bibitem{DeTar:1989kn}
C.~DeTar and T.~Kunihiro,
\newblock Phys.\ Rev.\ D \textbf{39} (1989), 2805.

\bibitem{Jido:1998yk}
D.~Jido, M.~Oka and A.~Hosaka,
\newblock Phys.\ Rev.\ Lett.\ \textbf{80} (1998), 448.

\bibitem{Jido:1998av}
D.~Jido, Y.~Nemoto, M.~Oka and A.~Hosaka,
\newblock Nucl.\ Phys.\ A \textbf{671} (2000), 471.

\bibitem{Jido:2001nt}
D.~Jido, M.~Oka and A.~Hosaka,
\newblock Prog.\ Theor.\ Phys.\ \textbf{106} (2001), 873.

\bibitem{Kaiser:1995cy}
N.~Kaiser, P.B.~Siegel and W.~Weise,
\newblock Phys.\ Lett.\ B \textbf{362} (1995), 23.

\bibitem{Inoue:2001ip}
T.~Inoue, E.~Oset and M.J.~Vicente~Vacas,
\newblock Phys.\ Rev.\ C \textbf{65} (2002), 035204.

\bibitem{Weinberg:1968de}
S.~Weinberg,
\newblock Phys.\ Rev.\ \textbf{166} (1968), 1568.

\bibitem{Weinberg:1979kz}
S.~Weinberg,
\newblock Physica A \textbf{96} (1979), 327.

\bibitem{Gasser:1984yg}
J.~Gasser and H.~Leutwyler,
\newblock Ann.\ Phys.\ \textbf{158} (1984), 142.

\bibitem{Cohen:1997sb}
T.D.~Cohen and X.D.~Ji,
\newblock Phys.\ Rev.\ D \textbf{55} (1997), 6870.

\bibitem{Ioffe:1981kw}
B.L.~Ioffe,
\newblock Nucl.\ Phys.\ B \textbf{188} (1981), 317.

\bibitem{Christos:1983kc}
G.A.~Christos,
\newblock Z.\ Phys.\ C \textbf{21} (1983), 83; ibid.\ \textbf{29} (1985), 361; 
Phys.\,Rev.\,D\,\textbf{35} (1987), 330.

\bibitem{Kim:1998up}
H.c.~Kim, D.~Jido and M.~Oka,
\newblock Nucl.\ Phys.\ A \textbf{640} (1998), 77.

\bibitem{LeeBook1972}
B.W.~Lee,
\newblock Chiral Dynamics (Gordon and Breach, 1972).

\bibitem{Jido:1999hd}
D.~Jido, T.~Hatsuda and T.~Kunihiro,
\newblock Phys.\ Rev.\ Lett.\ \textbf{84} (2000), 3252.

\bibitem{Jido:2002yb}
D.~Jido, H.~Nagahiro and S.~Hirenzaki,
\newblock Phys.\ Rev.\ C \textbf{66} (2002), 045202.

\bibitem{Jido:2002mg}
D.~Jido, H.~Nagahiro and S.~Hirenzaki,
\newblock Nucl.\ Phys.\ A \textbf{721} (2003), 665c.

\bibitem{Nagahiro:2003iv}
H.~Nagahiro, D.~Jido and S.~Hirenzaki,
\newblock nucl-th/0304068, to be published in Phys.\ Rev.\ C.

\bibitem{Hatsuda:1999kd}
T.~Hatsuda, T.~Kunihiro and H.~Shimizu,
\newblock Phys.\ Rev.\ Lett.\ \textbf{82} (1999), 2840.

\bibitem{Jido:2000bw}
D.~Jido, T.~Hatsuda and T.~Kunihiro,
\newblock Phys.\ Rev.\ D \textbf{63} (2001), 011901.

\bibitem{Waas:1997pe}
T.~Waas and W.~Weise,
\newblock Nucl.\ Phys.\ A \textbf{625} (1997), 287.

\bibitem{Inoue:2002xw}
T.~Inoue and E.~Oset,
\newblock Nucl.\ Phys.\ A \textbf{710} (2002), 354.

\bibitem{Jido:2000nt}
D.~Jido, M.~Oka and A.~Hosaka,
\newblock Prog.\ Theor.\ Phys.\ \textbf{106} (2001), 823.

\end{thebibliography}

\end{document}